\documentclass{PoS}
\usepackage{here}
\title{Energy dependence of nucleon-nucleon potentials in lattice QCD}
\ShortTitle{Energy dependence of NN potentials in lattice QCD}

\author{\speaker{Keiko Murano}\\
  Graduate School of Pure and Applied Sciences, University of Tsukuba,\\
  Tsukuba, Ibaraki 305--8577, Japan\\
  E-mail: \email{murano@het.ph.tsukuba.ac.jp}}
\author{Noriyoshi Ishii\\
  Department of Physics, University of Tokyo,\\
  Tokyo 113--0033, Japan\\
  E-mail: \email{ishii@ribf.riken.jp}}
\author{Sinya Aoki\\
  Graduate School of Pure and Applied Sciences, University of Tsukuba,\\
  Tsukuba, Ibaraki 305--8577, Japan\\
  E-mail: \email{saoki@het.ph.tsukuba.ac.jp}}
\author{Tetsuo Hatsuda\\
  Department of Physics, University of Tokyo,\\
  Tokyo 113--0033, Japan\\
   E-mail: \email{hatsuda@phys.s.u-tokyo.ac.jp}}
  \author{for HAL QCD Collaboration}
   \abstract{Recently a new approach to calculate the nuclear potential from
   lattice QCD has been proposed.
   In the approach the nuclear potential is constructed from Bethe-Salpeter (BS) wave
   functons through the Schr\"oedinger equation. The procedure leads to non-local but energy
  independent potential, which can be expanded in terms of local functions.
  In several recent applications of this method, local potentials, which correspond to
  the leading order (LO) terms of the expansion, are calculated from the
  BS wave function at $E\simeq 0$ MeV, where $E$ is the center of mass
  energy. It is therefore important to check the validity of the LO
  approximation obtained at $E\simeq 0$. 
 In this report, in order to check how well the LO approximation for the NN potentials works,
 we compare the LO potentials determined from the BS wave function at $E\simeq 45$ MeV  
 with those at $E\simeq 0$ MeV in quenched QCD.
 We find that the difference of the LO potentials between two energies
 are not found wihin the statistical errors.
  This shows that the LO approximation for the potential is valid at low energies to describe the NN interactions.   
     }
\FullConference{The XXVII International Symposium on Lattice Field Theory - LAT2009\\
		 July 26-31 2009\\
  		 Peking University, Beijing, China}

\newcommand{\vr}{\vec{r}}

\begin{document}
 \section{Introduction}

 The nucleon-nucleon (NN) potential is widely used in nuclear physics.
   Once the NN potential is known, one can, in principle, determine various structure of nuclei, by
   simply solving the corresponding Schr\"odinger equation.
   In the past few decades,  several different  NN potentials, such as
   phenomenological potentials determined from the fit of NN scattering data with $\chi^2/{\rm dof} \sim 1$ at $T_{\rm lab} < 300$
    MeV \cite{Machleidt:2000ge,Wiringa:1994wb,Stoks:1994wp}
    or  potentials based on effective field theory \cite{Weinberg:1990rz}, 
   have been proposed.
   Although these potentials can reproduce the scattering data to a quite
   good accuracy, they have a drawback that they need a
   large number of parameters to describe the phase shift of the NN scattering
   ($\mbox{AV}_{18}$ needs 40 and ChPT in $\mbox{N}^3$LO needs 24
   parameters. ). 
   In this respect, the NN potential from Lattice QCD proposed in
   Ref. \cite{Ishii:2006ec, Aoki:2008hh,
   Aoki:2009ji} has an advantage: 
   It requires  only a few fundamental parameters of QCD, the gauge
   coupling constant $g^2$ and quark masses $m_u, m_d, m_s,\cdots$, so
   that the method can be also applied to hyperon systems
   ($N\Lambda$,$N\Sigma$ \ldots) \cite{Nemura:2008sp}, for which only a
   limited number of experimental information is obtained so far.
 
 Recently  Hadrons to Atomic nuclei from Lattice (HAL) QCD collaboration is formed to study various aspect of baryon-baryon potentials based on the first principle of QCD. 
 In the approach by the HAL QCD collaboration, the non-local but energy-independent potential $U(r,r')$ is first constructed from
the Bethe-Salpeter (BS) wave function via the Schr\"odinger equation \cite{Ishii:2006ec, Aoki:2008hh,
   Aoki:2009ji}. 
   The non-local potential $U(r,r')$ can be described in terms of local functions using the derivative expansion, 
   whose leading order contribution gives the local potential.
   In several applications of this method, the leading order local potentials have been evaluated at
    zero energy in the center of mass frame.
    Therefore, in practice,  it is important to know at which energy range the leading order local potential is accurate enough to approximate the non-local potential  $U(r,r')$, which is faithful to the
scattering data by construction.
In this report, we extract the leading order local potentials at non-zero energy from the quenched lattice QCD simulation, and compare them with those previously obtained at zero energy.
A difference between them gives  an estimate of higher order correction in the derivative expansion. 

     This report is organized as follows.
     In section \ref{V-intro}, we give a brief review of the method to extract the NN potential in
     Lattice QCD using the derivative expansion.
         In section \ref{result}, we compare the leading local potentials between zero and non-zero energies. We have found that a difference between them is small compared to statistical errors.
 In section \ref{contamination}. we consider contaminations to the potentials from excited states, which become manifest at large separation where the potential is expected to vanish.    
          Section \ref{summary} is devoted to summary and conclusion.

 \section{NN potential from Lattice QCD}
 \label{V-intro}
 The non-local potential $U(\vec{r},\vec{r}^\prime)$ is constructed from
 the equal-time Bethe-Salpeter (BS) wave function $\phi(\vec{x},k)$
 through the Shr\"odinger equation \cite{Aoki:2008hh,Aoki:2009ji} as
 \begin{eqnarray}
  \left(\triangle + k^2\right) \ \phi(\vec{r};k) = m_N \int d^3 \ r^\prime \
   U(\vec{r},\vec{r}^\prime) \ \phi(\vec{r}^\prime,k), \label{eq:SE}
 \end{eqnarray}
 where "$k$" denotes the ``asymptotic momentum'', which
 is related to the total relativistic energy $W$ as
 $W=2\sqrt{m^2_N+k^2}$.
 The derivative expansion up to $\vec{\nabla}$, together with
various constraints from symmetries leads to the
 conventional form of the NN potential at low energies widely used in
 nuclear physics \cite{okubo}:
 \begin{eqnarray}
  U(\vec{r},\vec{r}^\prime) = \left[
			V_0^I(r) +
			V_\sigma^I(r) \ \left(\sigma_1\cdot\sigma_2\right) + V_T^I(r) \ S_{12} +
			V_{LS}^I(x) \ \vec{L}\cdot \vec{S} + \mathcal{O}(\vec{\nabla^2})
		       \right]\delta(\vec{r}-\vec{r}^\prime), \label{eq:delive}
 \end{eqnarray}
 where $r=\vert \vec r\vert$, $S_{12} =
3 (\vec{\sigma_1\cdot\vec{r}})(\vec{\sigma_2}\cdot\vec{r})/r^2-\vec{\sigma_1}
 \cdot \vec{\sigma_2}$ is the tensor operator, $\vec{S} = \left(\vec{\sigma_1} + \vec{\sigma_2}\right)/2$ is the total spin,  $\vec L =\vec r\times \vec p$ is the orbital angular momentum, and $I=0,1$ is the total isospin. Note that $\vec L\cdot\vec S$, which is $\mathcal{O}(\vec\nabla)$,  is of next-to-leading order in the expansion.
 Eq. (\ref{eq:SE}) with (\ref{eq:delive}) successively determines local functions $V^I_A(x)$ ($ A=0,\sigma, T, LS, \cdots$).

  The BS wave function on the lattice with the lattice size
  $L$ is defined by
  \begin{eqnarray}
 \phi^S(\vec r;k) = \frac{1}{L^3}\displaystyle\sum_{\vec{x}}P^{S}_{\alpha
  \beta}\langle 0 |
  \hat{n}_\beta\left(\vec{r}+\vec{x}\right)\hat{p}_\alpha(\vec{x})|B=2;W\rangle,
  \hspace{1cm} W = 2\sqrt{m_N^2+k^2},
  \label{eq:BSwave}
  \end{eqnarray} 
 where $W$ is the total energy of the two nucleon system in the center of mass system, and
 $P^{\sigma}_{\alpha\beta}$ denotes a projection operator to
 the spin singlet state
 ($P^{S=0}_{\alpha,\beta}=(\sigma_2)_{\alpha,\beta}$) or triplet
 state ($P^{S=1}_{\alpha,\beta}=(\sigma_1)_{\alpha,\beta}$) for spinor
 indices $\alpha$ and $\beta$.
 Local composite operators for the proton and the neutron
 $\hat{n}$ and $\hat{p}$ are given by 
 \begin{eqnarray}
  \hat{n}_\beta(y) &=&
   \epsilon_{abc}\left(\hat{u}_\alpha(y)C\gamma_5\hat{d}_b(y)\right)\hat{d}_{c\beta}(y), \hspace{1cm}
  \hat{p}_\alpha(x) =
   \epsilon_{abc}\left(\hat{u}_\alpha(x)C\gamma_5\hat{d}_b(x)\right)\hat{u}_{c\alpha}(x),\label{eq:operator}
 \end{eqnarray}
  where, $a,b,c$ denote color indices, and $C$ is the charge conjugation matrix.

In this report, we consider potentials for the $^1S_0$ state and the $^3S_1$ -- $^3D_1$ state.
In the case of $^1S_0$ state, the Schr\"odinger equation at leading order becomes
\begin{eqnarray}
(\triangle + k^2)\phi^{^1S_0}(r;k) = 2\mu V^{^1S_0}(r) \phi^{^1S_0}(r;k)
\end{eqnarray}
for the spin singlet channel ($S=0$) with the reduced mass $\mu = m_N/2$,
where the wave function  $\phi^{^1S_0}(r;k)$ for the $^1S_0$ state is given by the projection $P$ as
\begin{eqnarray}
\phi^{^1S_0}(r;k) &=& P \phi^{0}(\vec r;k) \equiv
\frac{1}{24} \displaystyle \sum_{R\in O} \phi^{0}(R[\vec r];k) .
 \end{eqnarray}
 Here the summation over $R\in O$ is taken for the cubic transformation group
 to project out the $A_1^+$ state.   Then  
 the central potential is easily  obtained as
  \begin{eqnarray}
   V^{^1S_0}(r)&\equiv&   V_0^1(r) - 3 V_\sigma^1 (r) = E + \frac{1}{2\mu}\frac{\triangle
    \phi^{^1S_0}(r;k)}{\phi^{^1S_0}(r;k)}, 
    \label{eq:local}
  \end{eqnarray}
  where $E(=\frac{k^2}{2\mu})$ is an effective kinetic energy in the center of mass system.
   
 For the spin triplet channel ($S=1$),  the Schr\"odinger equation at leading order becomes more complicated due to the mixing between $^3S_1$ and $^3D_1$ components by the tensor potential:
\begin{eqnarray}
(\triangle + k^2) \phi^{1} (\vec r;k) &=& 2\mu \{ V_c(r) + V_T(r) S_{12} \} \phi^{1} (\vec r;k)
 \end{eqnarray}
 where $V_c(r)\equiv   V_0^0(r) +  V_\sigma^0 (r)$. 
By  two projections $P$ and $Q\equiv 1-P$, the above Schr\"odinger equation is decomposed as
 \begin{eqnarray}
 \left(
 \begin{array}{cc}
 P \phi^{1} (\vec r;k) & P S_{12}  \phi^{1} (\vec r;k) \\
 Q \phi^{1} (\vec r;k) & Q S_{12}  \phi^{1} (\vec r;k) \\
 \end{array}
 \right)
 \left(
 \begin{array}{c}
 V_c(r) \\
 V_T(r) \\
 \end{array}
 \right) &=& \frac{\triangle +k^2}{2\mu}
 \left(
 \begin{array}{c}
  P \phi^{1} (\vec r;k) \\
  Q  \phi^{1} (\vec r;k) \\
 \end{array}
 \right),
 \end{eqnarray}
 from which $V_c(r)$ and $V_T(r)$ can be separately obtained\cite{Ishii:2009zr, Aoki:2009ji}.

  \section{Numerical Simulations and results}\label{result}
  \subsection{Lattice QCD setup}
  We employ the standard plaquette gauge action 
  on a $32^3 \times 48$ lattice with the $\beta=5.7$ for quenched gauge configurations. 
 Quark propagators are calculated by the Wilson quark action at
  $\kappa=0.1665$. 
This setup leads to the lattice spacing $a^{-1}=1.44(2)$ GeV 
($a \sim 0.137$ fm) from $m_\rho$, the spatial extension $L=32a \sim 4.4$ fm,
$m_\pi\sim 0.53$ GeV and $m_N \sim 1.33$ GeV \cite{Fukugita:1994ve}.
  Quenched gauge configurations are generated by the heatbath
  algorithm with overrelaxation.  Potentials are measured on configurations separated by 200
  sweep. 4000 configurations are accumulated to obtain results in this report.
  These calculations are performed on Blue Gene/L at KEK.
  
The  BS wave function is obtained from the four-point correlator of nucleon
  operators in the large $t$ region,
  \begin{eqnarray}
 G^{(4)}(\vec{x},\vec{y},t,t_0)
 &=&\langle 0 | \hat{n}_\beta(\vec{y},t)
    \hat{p}_\alpha(\vec{x},t)\bar{\mathcal{J}}_{pn}(t_0)|0\rangle 
   =
    \sum_n A_n \langle 0|
    \hat{n}_\beta(\vec{y})\hat{p}_\alpha(\vec{x})|B=2;W_n\rangle
    e^{-W_n(t-t_0)} \nonumber \\
    &\sim&  A_0 \langle 0|
    \hat{n}_\beta(\vec{y})\hat{p}_\alpha(\vec{x})|B=2;W_0\rangle
    e^{-W_0(t-t_0)},
    \hspace{1cm} W_n = 2\sqrt{m_N^2+k_n^2}.
    \label{eq:4-point} 
  \end{eqnarray}
  Here   the source located at $t=t_0$, $\bar{\mathcal{J}}_{pn}(t_0)$, is defined by
  \begin{eqnarray}
   \bar{\mathcal{J}}_{pn} =  P^S_{\alpha^\prime \beta^\prime}
    \bar{P}_{\alpha'}(t_0)\bar{N}_{\beta'}(t_0), \hspace{1cm}
   \bar{P}_\alpha \equiv
    \epsilon_{a,b,c}\left(\bar{U}_aC\gamma_5\bar{D}_b\right)\bar{U}_{c\alpha}, \hspace{0.5cm}
   \bar{N}_\beta \equiv
    \epsilon_{a,b,c}\left(\bar{U}_aC\gamma_5\bar{D}_b\right)\bar{D}_{c\beta},\nonumber
  \end{eqnarray}
  where 
  \begin{eqnarray}
   U(t) &=& \sum_{\vec{x}} u(t,\vec{x})f(\vec{x}), \hspace{1cm}
   D(t) = \sum_{\vec{x}} d(t,\vec{x})f(\vec{x}),
  \end{eqnarray}
  where $f(x)$ is source function, as will be seen later.
  By examining the $t$ dependence of potentials, we see that ground state saturations for potentials are achieved at $t-t_0=9$. 
 \subsection{Periodic and anti periodic boundary conditions}
  The periodic boundary condition (PBC) is imposed on the quark fields along
  the spatial directions to obtain the NN potential at
  $E \sim 0$ MeV, while the anti-periodic boundary condition (APBC) is employed for the NN
  potential at $E\sim 3 \times (\pi/L)^2/m_N$, which corresponds to $45$ MeV in the center of mass system.
  For the PBC, we employ the wall source, i.e., $f(\vec{x}) = 1$ ,
  which enhance ground state of PBC, $\vec{p}=(0,0,0) \pi/L$.
  On the other hand,
  for the APBC, we employ four types of momentum wall sources,
  $f(\vec{x}) = \cos((+x+y+z)\pi/L)$,
  $\cos((-x+y+z)\pi/L)$,
  $\cos((+x-y+z)\pi/L)$,
  and $\cos((-x-y+z)\pi/L)$,
  where these sources enhance the ground state of
  APBC, i.e., $\vec{p}=(1,1,1) \pi/L$ state.
  Here, we have imposed positive parity to the system  by using the wall source
  with a cosine type instead of
  an exponential type.
  After the summation over the results of  four sources,
  the $A_1^+$ representation is obtained.
  To improve the statistics for the PBC case, we locate four sources on different time slices
  on each configuration.

To evaluate the central potential by eq. (\ref{eq:local}), we need to determine
  the first term, $k^2/2\mu$, which, 
  in principle, is obtained through the relation $W_0 =
  2\sqrt{m_N^2 + k^2}$.
  Within statistical and systematic errors, however, the values of
    $k^2$ turn out to be close to the free values.
  We therefore adopt the free values,  $E=k^2/m_N=0$ MeV (PBC) and
  $E=k^2/m_N=3(\pi/L)^2/m_N= 45$ MeV (APBC), to calculate potentials in this report.

\subsection{Comparison of local potentials between two energies}
 \begin{figure}[tb]
   \begin{tabular}{ccc}
    \scalebox{0.36}{\includegraphics{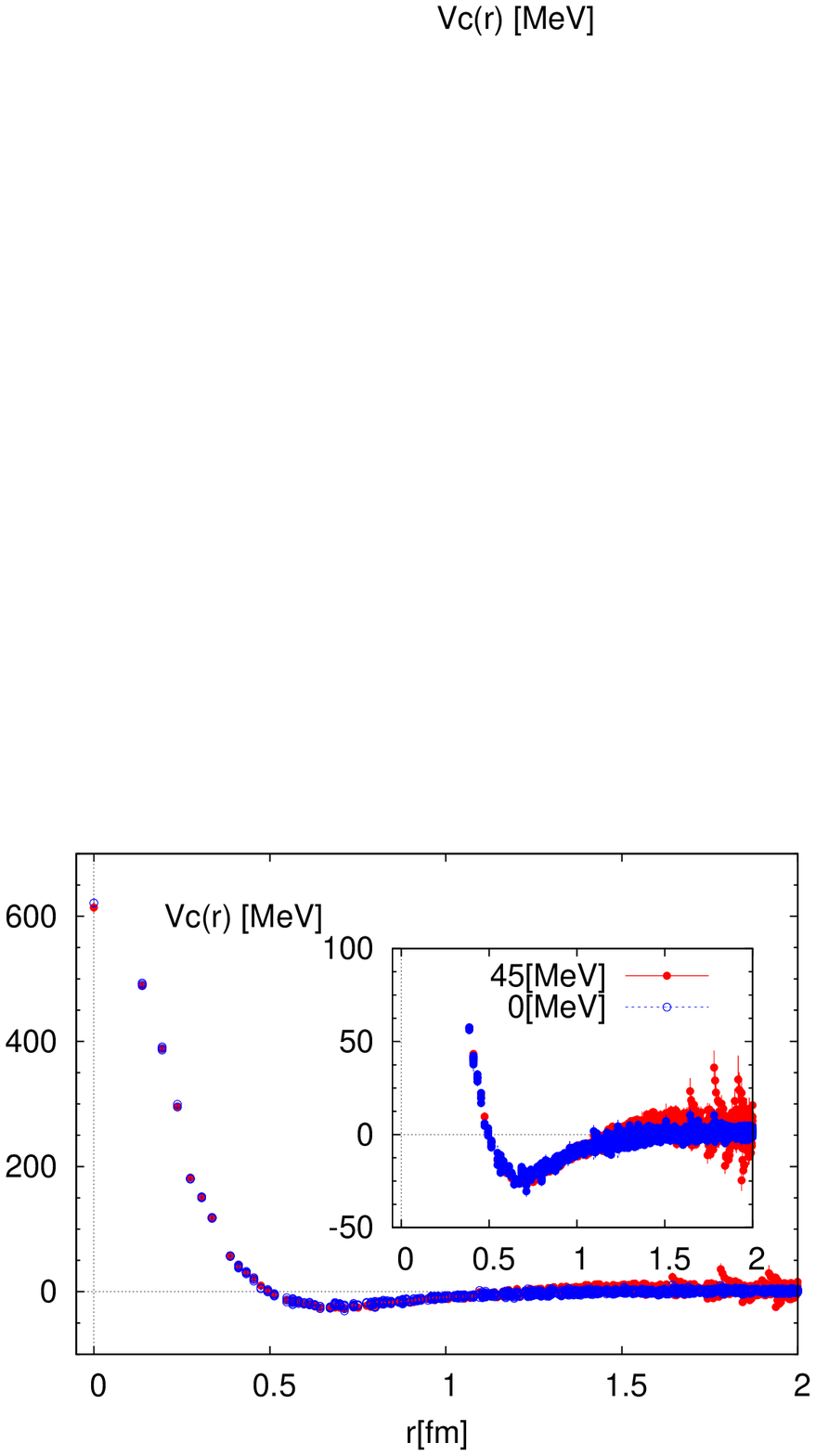}}
    &
    \scalebox{0.36}{\includegraphics{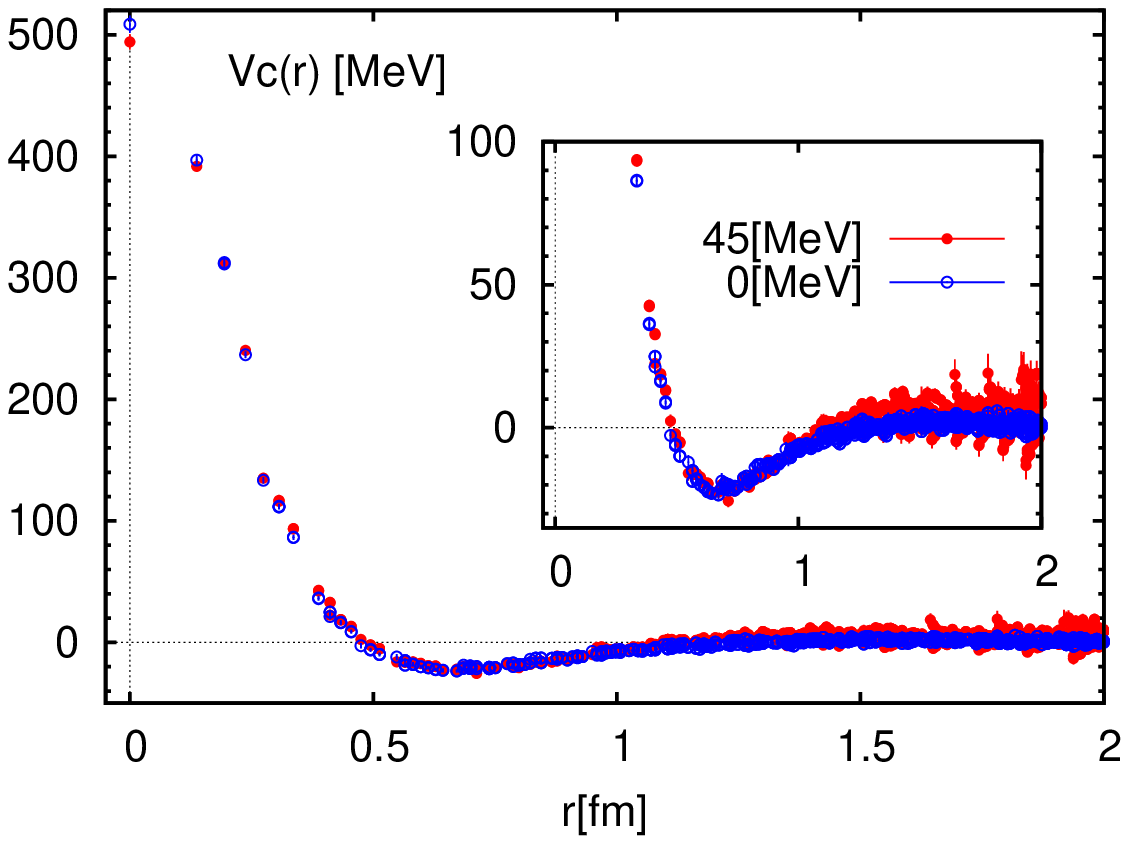}}
    &
    \scalebox{0.36}{\includegraphics{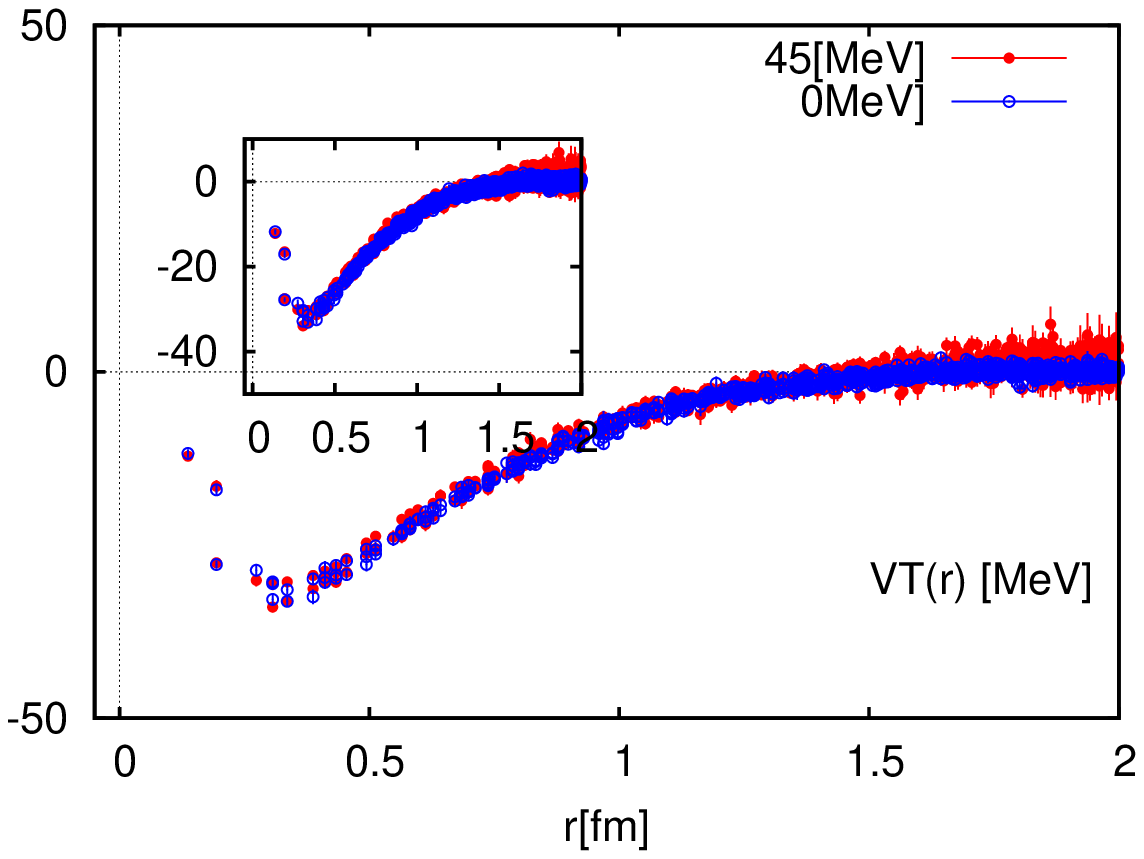}}
   \end{tabular}
  \caption{Comparison of local potentials with the PBC (blue open
  circle) and the  APBC (red closed circle) at $t-t_0=9$ and $r \le 2.0$ fm. The left  figure shows 
  the $^1S_0$ central potential, while the center and the right represent
  the $^3S_1-^3D_1$ central and tensor potentials, respectively.} 
  \label{fig:APBC_vs_PBC}
 \end{figure}
In Figure \ref{fig:APBC_vs_PBC}, we compare potentials at the leading
order of the derivative expansion obtained at $E\simeq 45$ MeV (red circles) with those at $E\simeq 0$ MeV (blue circles),
for the $V^{^1S_0}(r)$ (the left), $V_c(r)$ (the center) and $V_T(r)$ (the right). All data are taken at $t-t_0=9$, where grand state saturations for the potentials are achieved.  
From these figures we observe that the agreement of potentials between two energies
is quite good for all cases within the statistical errors.
We therefore conclude that the leading order contribution in the derivative expansion
gives a very accurate  approximation for the energy-independent non-local potential  $U(\vec r, \vec r^\prime)$
in the energy range from $E=0$ MeV to $E= 45$ MeV, in the case of the quenched approximation
at $m_\pi\simeq 530$ MeV.
Note that  the NLO contribution is absent for the spin-singlet  channel
while the NLO $V_{LS}$ potential exists for the spin-triplet.  The
agreement of $V_c$ and $V_T$ between two energies suggests that $V_{LS}
(r)$ is sufficiently small below $E=45$ MeV, at least for
$m_{\pi}\simeq 530$ MeV.
 \section{Contamination from excited states at large distances}
 \label{contamination}

\begin{figure}[here]
  \begin{tabular}{ccc}
   \scalebox{0.36}{\includegraphics{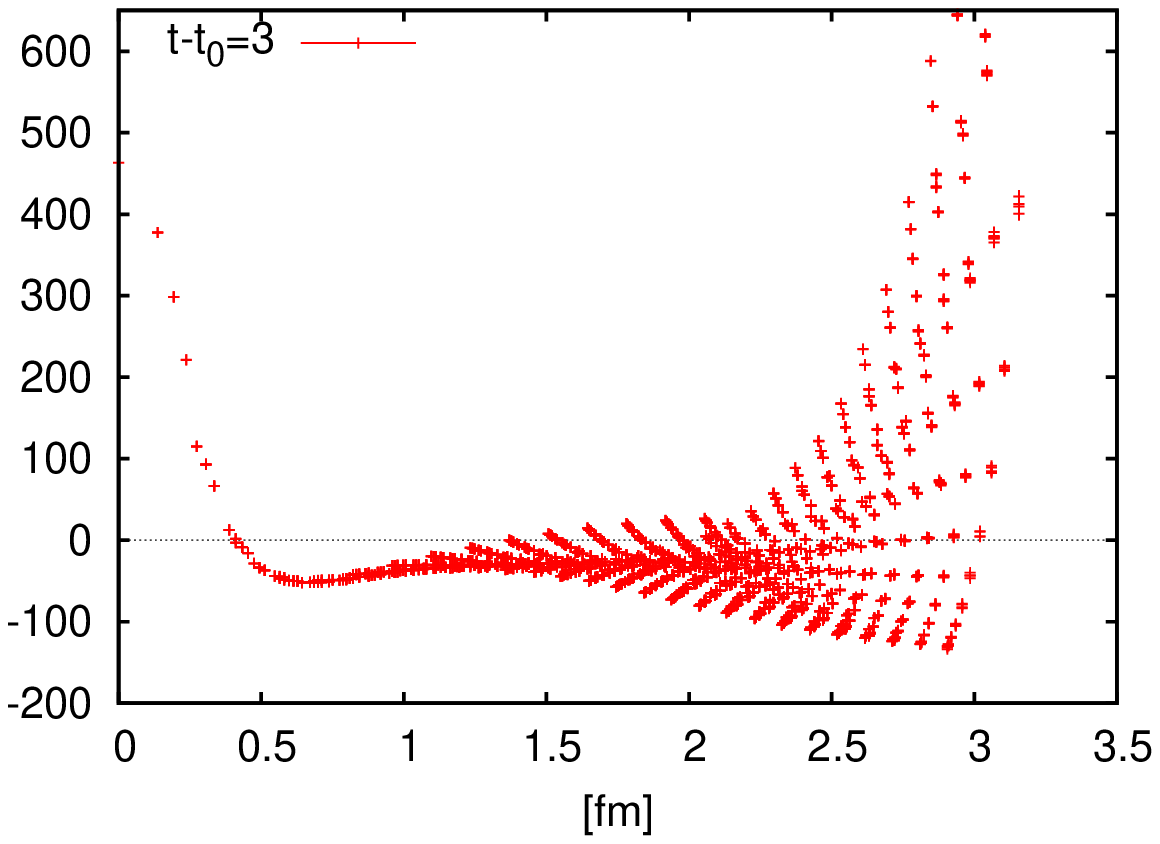}}
   &
   \scalebox{0.36}{\includegraphics{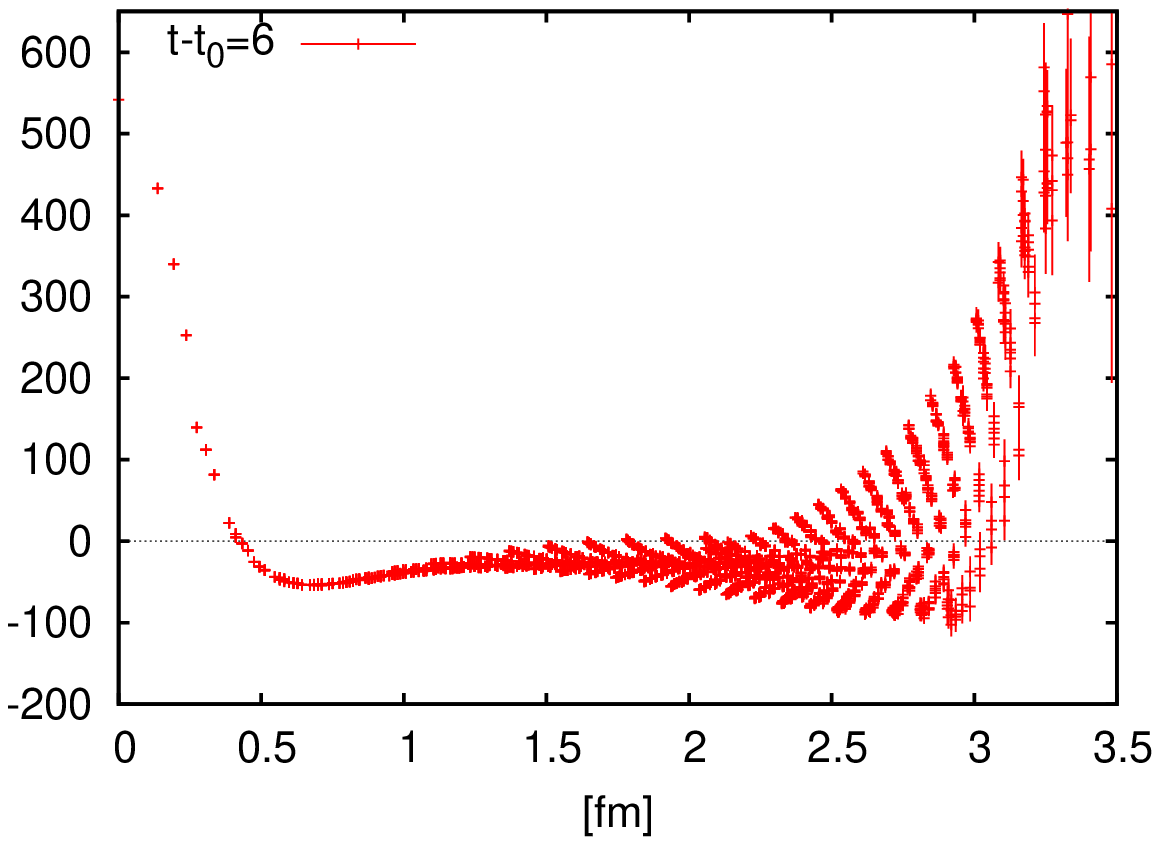}}
   &
   \scalebox{0.36}{\includegraphics{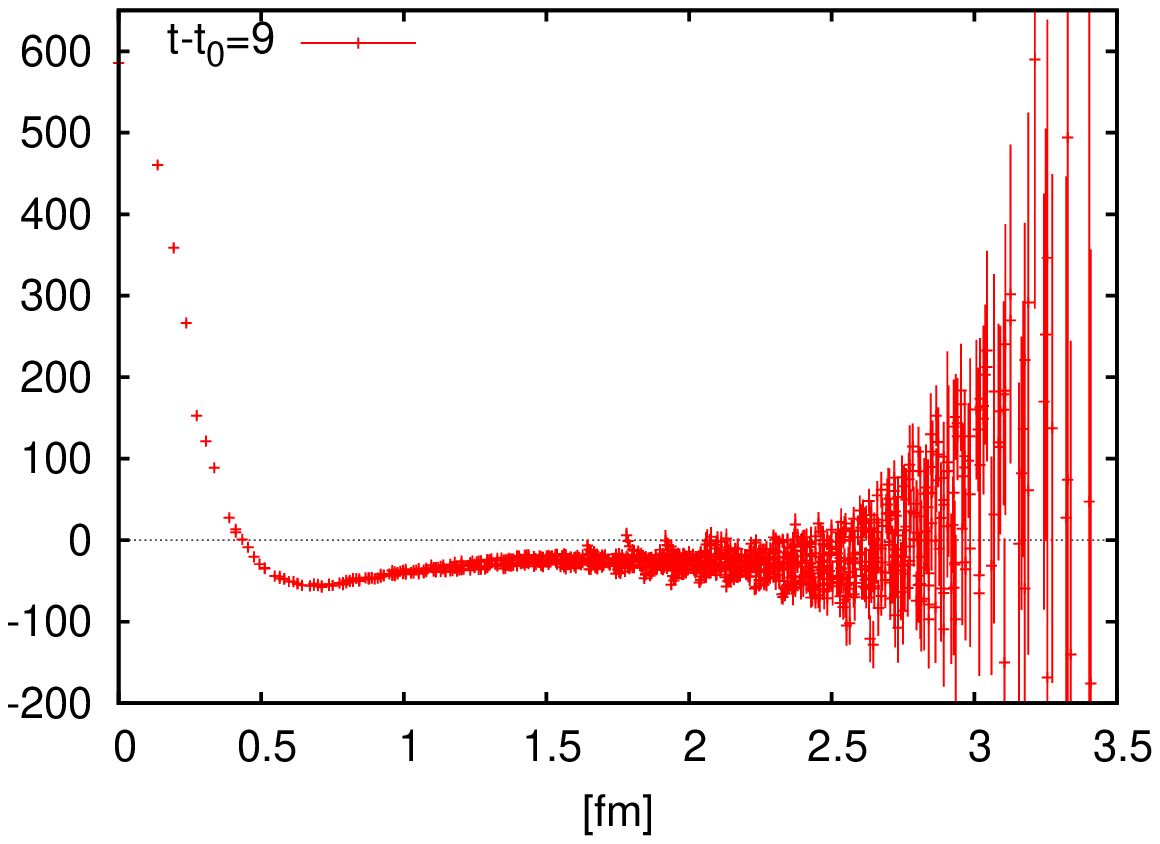}}
  \end{tabular}
  \caption{These figures are $^1S_0$ central potential with the APBC at
  $t-t_0 = 3$ (left), $6$ (center) and $9$ (right). These potential
  deviate from zero at $r > 2$ fm. Deviations of potentials are
  decrease as $t-t_0$ increases.}
  \label{fig:APBC_t-dep}
 \end{figure}

 Results in the previous section show that the local potentials obtained from the BS wave function at $E\simeq 45$ MeV agree well with those at $E\simeq 0$ MeV.
 One may notice, however, that potentials obtained at $E\simeq 45$ MeV (APBC) deviate from zero at
 large $r$, where potentials are expected to vanish. These deviations are not statistical fluctuations , as
 seen in Fig.\ref{fig:APBC_t-dep}, where the local potentials obtained with the APBC are plotted as a function of $r$  at $t-t_0 = 3, 6, 9$: Deviations of the potentials from zero at large distances have characteristic structures, which are most clearly seen at $t-t_0=3$, and the deviations
 decrease as  $t-t_0 $ increases.
 These observations suggests that the deviations of the potential from zero are caused by contaminations of excited states. 

Here we discuss how contaminations of excited states to the BS wave function affects the form of the corresponding potential. 
Let us assume that the BS wave function extracted from the 4 point function is dominated by 
the grand state with a small contamination of one excited state as follows.
\begin{eqnarray}
 \phi(\vec r,t) = \psi(\vec r, k_0) \ e^{-W_0 \ t} + \psi(\vec r,  k_1) \ e^{-W_1 \ t}, \hspace{1cm}
  (k_0 < k_1),
\label{eq:mix}
\end{eqnarray}
where $W_n = 2\sqrt{m_N^2+k_n^2}$, and $\psi(\vec r,k_0)$ is the BS wave function of the grand state with
 $E_0 = k_0^2/m_N$
 while $\psi(\vec r,k_1)$  is the wave function of the excited state
 with $E_1 = k_1^2/m_N$.
At sufficiently large $r$,  we assume that both wave function satisfy
\begin{eqnarray}
  \frac{1}{m_N}(\Delta + k_0^2) \ \psi(\vec r, k_0) =  0 \\
 \frac{1}{m_N}(\Delta + k_1^2) \ \psi(\vec r, k_1) =   0 .
 \label{eq:k1}
\end{eqnarray}
By inserting eq. (\ref{eq:mix})  into eq. (\ref{eq:local}), we arrive at the following expression:
  \begin{eqnarray}
   V_{\rm mix}(\vec r) &=& \frac{1}{m_N}\frac{\Delta \phi(\vec r,t)}{\phi(\vec r,t)} +\frac{k_0^2}{m_N}
    \stackrel{r\rightarrow\infty}{\longrightarrow} 
    \frac{1}{m_N}\left(k^2_1-k^2_0\right) \ P(\vr), \label{eq:model}
  \end{eqnarray}
  where $P(\vr)$ is given by
  \begin{eqnarray}
 P(\vr) = \frac{\psi(\vr, k_1)}{\psi(\vr, k_0)+\psi(\vr, k_1)e^{-t (W_1-W_0)}}e^{-t  (W_1-W_0)}  .
  \end{eqnarray}
We see that, even in the non-interacting region where the true potential $V(\vec r) $ vanishes, eq
(\ref{eq:model}) leads to a non-vanishing "potential"  as
\begin{eqnarray}
 V_{\rm mix}(\vec r) = \frac{k_1^2-k_0^2}{m_N} P(\vr) \neq 0.
\end{eqnarray}

The agreement of local potentials between two energies  indicates that
the effect of these contaminations to potentials is smaller than statistical errors at short distance ($r< 2.0$ fm). Therefore the conclusion in the previous section remains true.
Since the true potentials vanish at long distance, however, the small effect  due to the contaminations can become significant. Note that  the APBC implies that not only the numerator but also the denominator of $P(\vr)$ vanish at boundaries, so that $P(\vr)$ could become large near boundaries.
The consideration so far suggests that the deviations of the potentials from zero with the APBC at large $r$ is caused by the contaminations from the excited states.  
Our choices of momentum wall sources creates not only the grand state
with $\vec p = \frac{\pi}{L}(\pm1,\pm1,\pm1)$  but also the excited
state with  $\vec p = \frac{\pi}{L}(\pm3,\pm3,\pm3)$. Moreover the
energy difference in lattice unit is not so large: $W_1-W_0 \simeq 0.23/a$ ($\sim 360$ MeV) in this case. Therefore it is likely that the contamination comes from the $(3,3,3)$ excited state.  
This observation, however, needs to be confirmed, which is now underway.
\section{Summary and conclusion}
\label{summary}
We have examined how well the leading order contributions in the derivative expansion of the non-local potential $U(r,r')$ describe the NN interactions in the wide range of energy.
We have compared the local NN potentials for the $^1S_0$ state (the
central potential) and for the $^3S_1$--$^3D_1$ state (the central and
the tensor potentials) obtained at $\simeq 45$ MeV
with those obtained at $E\simeq 0$ MeV in quenched QCD at $m_\pi \simeq 530$ MeV.
We have found that differences of these potentials between two energies are very small. 
From this result we conclude that the leading order local potentials in the derivative expansion 
are good approximations for the NN potentials at energy up to 45 MeV in quenched QCD, and
that the local potential constructed at $E\sim 0$ MeV can be used to investigate  properties of the NN interaction at low energy.

In the future it is important to apply the analysis in this report 
to the NN potentials in full QCD at lighter pion mass and to more general potentials including hyperons,
in order to confirm the validity of the leading order local potential approximation for these cases.

 \section*{Acknoendwledgments}
 We are grateful for authors and maintainers of {\tt CPS++}\cite{cps}, of
 which a modified version is used for simulations done in this work. 
 This work is supported by the Large Scale Simulation Program
 No.08-19(FY2008) and
 No.09-23(FY2009) of High Energy Accelerator Research Organization (KEK).
 This work was supported in part by the Grant-in-Aid of the Ministry of Education, Science and Technology, Sports and Culture (Nos. 20340047, 20105001, 20105003).

\end{document}